\documentclass[12pt]{article}
\usepackage{amsfonts}
\usepackage{amsmath,mathrsfs,bm,amssymb,color}
\usepackage[width=15cm,height=8.5in]{geometry}
\usepackage[dvipdfmx]{hyperref}

\newtheorem{theorem}{Theorem}[section]
\newtheorem{definition}[theorem]{Definition}
\newtheorem{lemma}[theorem]{Lemma}
\newtheorem{proposition}[theorem]{Proposition}
\newtheorem{corollary}[theorem]{Corollary}
\newtheorem{remark}[theorem]{Remark}

\providecommand{\bc}[1]{\begin{corollary}\label{#1}}
\providecommand{\ec}{\end{corollary}}
\providecommand{\bp}[1]{\begin{proposition}\label{#1}}
\providecommand{\ep}{\end{proposition}}

\newenvironment{proof}[1][Proof]{{\sc #1.} }{\hfill $\Box$}

\newcommand{\ren}{E_{\rm ren}^\Lambda}

\newcommand{\bt}[1]{\begin{theorem}\label{#1}}
\newcommand{\et}{\end{theorem}}
\newcommand{\bl}[1]{\begin{lemma}\label{#1}}
\newcommand{\el}{\end{lemma}}

\newcommand{\br}[1]{\begin{remark}\label{#1}}
\newcommand{\er}{\end{remark}}

\makeatletter
\@addtoreset{equation}{section}
\makeatother

\def\bbbone{{\mathchoice {\rm 1\mskip-4mu l} {\rm 1\mskip-4mu l}
{\rm 1\mskip-4.5mu l} {\rm 1\mskip-5mu l}}}
\def\one{\bbbone}

\newcommand{\akl}{A_{\Lambda}}

\renewcommand{\k}{\kappa}

\newcommand{\vz}{{\hat {\varrho}_0}}
\newcommand{\liml}{\lim_{\la\rightarrow \infty}}

\newcommand{\ai}{A_{\infty}} 
\newcommand{\cp}{C_{\rm p}} 
\newcommand{\am}{C_\ast}
\newcommand{\suk}{\sup_{|k|\geq 1}} 
 
\newcommand{\GGG}{(H_\La-\El+\omega(k))^{-\han}} 
\newcommand{\GGGG}{(H_\La-\El+\omega(k))\f}
\newcommand{\FF}{(H_0+ \one)^{-\han}} 

\newcommand{\grr}{\Psi_{\la'}}
\newcommand{\gz}{H_0}
\newcommand{\ckl}{c_{\la}} 
\newcommand{\vkl}{V_{\la}} 
\newcommand{\HL}{H_\La}

\newcommand{\El}{E_{\la}} 
\newcommand{\eel}{E_{\la}^\ast} 
\newcommand{\Ei}{E_{\infty}}
\newcommand{\at}{g^2}
\newcommand{\mm}{|g|} 

\providecommand{\FFF}{(H_0+\one)^{1/2}}
\newcommand{\m}{\left\|{\vpl}/{\sqrt\omega}\right\|}
\providecommand{\epp}{E_{\rm p}}
\newcommand{\pa}{H_{\rm p}}
\newcommand{\eht}{e^{-t\HL }}

\newcommand{\kp}{\one_{|k|\geq\la}}

\providecommand{\pro}[1]{(#1_t)_{t\geq 0}}
\newcommand{\Ebb}{{\mathbb E}}

\providecommand{\eq}[1]{\begin{equation}\label{#1}}
\providecommand{\en}{\end{equation}}

\renewcommand{\d}{\displaystyle}

\providecommand{\qed}{\hfill {$\Box$}\par\medskip}
\providecommand{\BR}{{\mathbb{R}^3 }}

\providecommand{\bi}
{\begin{itemize}}
\providecommand{\ei}{\end{itemize} }
\providecommand{\CC}{{\mathbb{C}}}
\providecommand{\RR}{\mathbb{R}}

\providecommand{\kak}[1]{(\ref{#1})}
\providecommand{\LR}{{L^2(\BR)}}

\providecommand{\hni}{H_{\rm ren}(P)}
\providecommand{\D}{{\ms D}}

\providecommand{\hn}{H(P)}

\providecommand{\ms}[1]{\mathscr{#1}}

\providecommand{\fff}{\mathscr{F}}
\providecommand{\ffff}{{\mathscr{F}}}

\providecommand{\is}{\inf \s}

\renewcommand{\sp}{\Sigma_{\rm p}}

\providecommand{\f}{^{-1}}
\providecommand{\lk}{\left(}
\providecommand{\rk}{\right)}
\providecommand{\lkk}{\left\{}
\providecommand{\rkk}{\right\}}

\providecommand{\add}{a^\ast}

\providecommand{\ov}[1]{\overline{#1}}
\providecommand{\hf}{H_{\rm f}}
\providecommand{\pf}{{P_{\rm f}}}

\providecommand{\gr}{\varphi_{\rm g}}
\providecommand{\grt}{\varphi_{\rm g}^T}
\providecommand{\JJ}{{\rm J}}
\providecommand{\gre}{\varphi_{\rm ren}}

\providecommand{\half}{\frac{1}{2}}
\providecommand{\han}{{1/2}}

\providecommand{\he}{H_\eps(P)}

\providecommand{\hi}{H_{\rm I}}
\providecommand{\hp}{H_{\rm p}}

\providecommand{\la}{{\lambda}}
\providecommand{\La}{{\Lambda}}

\providecommand{\s}{\sigma}
\providecommand{\hhh}{\ms H} 
\newcommand{\TTT}[1]{{\rm (#1)}}

\providecommand{\slim}{{\rm s}\!-\!\lim}

\providecommand{\vp}{{\hat  \varrho}}
\providecommand{\vpl}{{\hat  \varrho_\Lambda}}
\providecommand{\vpe}{{\hat  \varrho_\eps}}

\providecommand{\non}{\nonumber}

\providecommand{\vp}{\mathop{\mathrm{{\varphi}}}\nolimits}

\providecommand{\e}{e}

\makeatletter
\@addtoreset{equation}{section}
\makeatother

\providecommand {\D}  {\ensuremath {\mathcal{D}}}
\providecommand {\s}  {\ensuremath {\mathcal{S}}}

\providecommand{\eps}{\varepsilon}

\newcounter {constant}
\newenvironment{constant}{\refstepcounter{constant} }{}

\makeatletter
\@addtoreset{equation}{section}
\makeatother
\title
{\sc Translation invariant models in QFT 
 without ultraviolet cutoffs}
\author{Fumio Hiroshima\thanks{Faculty of Mathematics, Kyushu University, Fukuoka, Japan} }
\date{\today}
\begin{document}

\maketitle

\begin{abstract}
The translation invariant model in quantum field theory is considered by functional integrations.  
Ultraviolet renormalization of the translation invariant Nelson model with a  fixed total momentum  
is  proven by functional integrations. 
As a corollary it can be shown that the Nelson Hamiltonian with zero total momentum has a ground state for arbitrary values of coupling constants in two dimension. Furthermore the ultraviolet renormalization of the polaron model is also studied.   
\end{abstract}

\bigskip
\section{Introduction}
\subsection{The Nelson model with a fixed total momentum}
In this paper we consider 
an ultraviolet (UV) renormalization of 
the Nelson model  $H(P)$ with 
a fixed total momentum $P\in\BR$ by functional integrations. 

The Nelson model describes an interaction system  
between a scalar bose field and particles governed by 
a Schr\"odinger operator.  
The interaction is linear in a field operator and the model is one of a prototype of  interaction models in quantum field theory. 
The Nelson Hamiltonian can be realized as a self-adjoint operator $H$ on a Hilbert space and the spectrum of 
$H$ has been studied so far from several point of view. 
See Appendix \ref{A} for the Nelson model. 
In the case where external potential is dropped  in 
$H$, the Hamiltonian turns to be translation invariant, 
and it  can be  realized as the family of self-adjoint operators $H(P)$ indexed by the so-called total momentum $P\in\BR$. 
The spectrum of $H(P)$ 
is studied for every $P\in\BR$, and  
the difference of spectral property of $H(P)$ from every $P$ 
is interesting.  

Before giving the definition of $H(P)$, we prepare 
tools  used in this paper. 
The boson Fock space
 $\fff$
over 
$\LR$ is defined by 
\eq{fff}
\fff=\bigoplus_{n=0}^\infty [\otimes_s^n \LR].
\en
Here 
$\otimes_s^n \LR$ describes $n$ fold symmetric tensor product of $\LR$ with $\otimes_s^0 \LR=\CC$. 
Let $\add (f)$ and $a(f)$, $f\in\LR$, be the creation operator and the annihilation operator, respectively, in $\fff$, which 
satisfy  canonical commutation relations:  
$$[a(f),\add(g)]=(\bar f, g),\quad
[a(f),a(g)]=0=[\add(f), \add(g)].$$
Note that 
Here $(f,g)$ denotes the scalar product on $\LR$ and it is linear in $g$ and anti-linear in $f$. 
We also note that 
$f\mapsto \add(f)$ and $f\mapsto a(f)$ are  linear. 
Denote the dispersion relation by $\omega(k)=|k|$ . 
Then the  free field Hamiltonian 
$\hf$ of $\fff$ is then defined by the second quantization of 
$\omega$,
i.e., $ \hf =d\Gamma(\omega)$, 
and the field momentum operator $\pf$ 
 by 
 $ \pf_\mu =d\Gamma(k_\mu)$ and we set 
 $\pf=(\pf_1,\pf_2,\pf_3)$.  
They satisfy 
\eq{sasa}
e^{-it\hf} \add(f) e^{-it\hf}=\add(e^{-it\omega}f),\quad 
e^{-it\hf} a(f) e^{-it\hf}=a(e^{it\omega}f).
\en
The field operator  $\phi=\phi(\vp)$ 
on which UV cutoff $\vp$ is imposed is defined by 
\begin{equation}
\label{hint}
\phi  = \frac{1}{\sqrt 2}\lk
\add(\vp/\sqrt{\omega})+
a(\widetilde{\vp}/\sqrt{\omega}) 
\rk. 
\end{equation}
Here $\vp$ denotes the Fourier transform of 
a cutoff function $\varrho$ satisfying 
 $\vp/\sqrt\omega\in\LR$, and $\widetilde{\vp}(k)=\vp(-k)$.

\begin{definition}
\TTT{Translation 
invariant  Nelson Hamiltonian}
 $H(P)$ 
is a linear  operator on $\fff$ and is  defined by 
\eq{Ham}
\hn = \half (P-\pf)^2 + \hf  + g\phi, \quad P\in \BR.
\en
\end{definition}
Before going to discussion on $H(P)$ we have to mention the self-adjointness of $H(P)$. 
We decompose $H(P)$ as 
$H_0+\hi(P)$ to show the self-adjointness, 
where 
\begin{align*}
&H_0=\half\pf^2+\hf,\\
&\hi(P)=\half |P|^2-P\cdot \pf +g \phi.
\end{align*}
Under the assumptions
\eq{sa}
{\vp }/{\sqrt \omega}\in \LR,\quad  {\vp }/{\omega}  \in   \LR ,\quad \ov{\vp(k)}=\vp(-k)
\en
we see that  
$\|\phi F\|\leq (1/\sqrt2) (2\|\vp/\omega\| \|\hf^\han  F\|+
\|\vp/\sqrt\omega \|\|F\|)
$ follows for $F\in D(\hf)$ and 
$\phi$ is symmetric. 
Then 
the interaction 
$\hi(P)$ is well defined, symmetric and it is 
infinitesimally 
$H_0$-bounded, i.e., for arbitrary $\eps>0$, 
there exists a $b_\eps>0$ such that 
$$\|\hi \Phi\|
\leq 
\eps  \|H_0\Phi\|+
b_\eps\|\Phi\|$$
for all $\Phi\in D(H_0)$. 
Thus
by the Kato-Rellich theorem $\hn$ is 
self-adjoint on $D(H_0)$ for every $P\in\BR$. 
Throughout this paper we assume condition \kak{sa}.

The purpose of this paper is to show 
UV  renormalization (=the point charge limit) of $\hn$. 
It is remarked that the point charge limit, $\vp\to \one$,  of $H(P)$ can be actually achieved in a similar manner to 
\cite{nel64} by functional analysis. 
While the purpose of this paper is
 to prove 
the point charge limit  by functional integrations. 
Machinery used in this paper 
is similar to  \cite{ghl13}, 
where it  plays an important role that 
 $e^{-tH}$ is positivity improving. 
Semi-group  $e^{-t\hn}$ is, however,  not positivity improving for $P\not=0$. Despite this fact we can achieve UV renormalization by using a diamagnetic inequality derived 
from functional integration.  

This paper is organized as follows. 
In Section 2 we show UV renormalization.
Section 3 is devoted to 
showing the existence of a renormalized ground state. 
In Section 4 we consider the polaron model. 
In Appendix we briefly introduce euclidean quantum field theory and the Nelson model.

\section{Renormalization}
\subsection{UV renormalization and main result}
Let $\la>0$ be an infrared cutoff parameter 
and we fix it throughout.  
Consider the cutoff function
\eq{sasa3}
\vpe(k)=\e^{-\eps|k|^2/2}\kp,\quad \eps>0
\en
and define the regularized Hamiltonian
by \begin{equation}
\he  = \half(P-\pf) ^2 + \hf  + 
g \phi_\eps,\quad \eps>0,
\end{equation}
where 
$
\phi_\eps$ is 
defined 
by $\phi$ with $\vp$ replaced by 
$\vpe$.
Here $\eps>0$ is regarded as the UV cutoff parameter. 
We investigate  the limit of $\he $ as 
$\eps\downarrow 0$. 
Precisely we can show the existence of a self-adjoint operator 
$H_{\rm ren}(P)$ such that 
\eq{cp}
\d e^{-T(H_\eps(P)-E_\eps)}\to e^{-T H_{\rm ren}(P)}
\en
by functional integrations, where 
\eq{sasa6}
E_\eps=-{g^2} \int_{|k|>\la}
\frac{\e^{-\eps|k|^2}}{2\omega(k)}\beta(k) dk
\en
denotes the renormalization term and  the propagator $\beta$ is given by 
\eq{beta}
\beta(k)=\frac{1}{\omega(k)+|k|^2/2}.
\en
Notice  that $E_\eps\to-\infty$ as $\eps\downarrow 0$.
Our main theorem shows  \kak{cp}
 for all $P\in\BR$.
\bt{main}
\TTT{UV renormalization}
Let $P\in \BR$. 
Then 
there exists a self-adjoint operator $\hni$ such that
\eq{main1}
\slim_{\eps\downarrow0}
\e^{-T(\he -E_\eps)}=\e^{-T\hni},\quad T\geq0.
\en
\et
We carry out the proof by functional integration and obtain $E_\eps$ as the diagonal term of a
pair interaction potential on the paths of a Brownian motion.

\subsection{Feynman-Kac type formula}
A Feynman-Kac type formula of 
 $(F,\e^{-T\he }G)$ is constructed  for 
 $F, G\in\fff$ and $P\in \BR$. 
Denote
\eq{yoshida}
H_{-k}(\RR^n)=\{f\in \ms S_\RR'(\RR^n)|\hat  f\in L_{\rm loc}^1(\RR^n), 
\omega^{-k/2}\hat  f\in
L^2(\RR^n)\}
\en
endowed with the norm $\d \|f\|_{H_{-k}(\RR^n)}^2=\int_{\RR^n} |\hat  f(x)|^2|x|^{-k} dx$. 
Recall that a
Euclidean field is a family of Gaussian random variables $\{\phi_{\rm E}(F),  F\in H_{-1}(\RR^4)\}$ on a
probability space $(Q_{\rm E}, \Sigma_{\rm E}, \mu_{\rm E})$, such that the map $F \mapsto \phi_{\rm E}(F)$
is linear, and their mean and covariance are given by
$$
\Ebb_{\mu_{\rm E}}[\phi_{\rm E}(F)]=0 \quad \mbox{and} \quad \Ebb_{\mu_{\rm E}}[\phi_{\rm E}(F)
\phi_{\rm E}(G)]=\half (F,G)_{H_{-1}(\RR^4)}.
$$
See Appendix \ref{appb} for the detail. 
Let $(B_t)_{t\in\RR}$ be the $3$-dimensional Brownian motion on the hole real line on the Wiener space. Let $\Ebb[\cdots]$ be the expectation with respect to the Wiener measure starting from zero. 
\bl{FKF}
\TTT{Feynman-Kac type formula}
Let $F, G\in \fff$. Then 
it follows that 
\begin{align}
(F, \e^{-2TH_\eps(P)}G)
\label{fkf}
=  \Ebb \left[  
\lk 
\JJ_{-T}\e^{i(P-\hat {\pf})B_{-T}}
 F, 
\e^{-\phi_{\rm E}(\int_{-T}^T
 \delta_s\otimes \tilde\varrho_\eps(\cdot-B_s) ds)} \JJ_T
 \e^{i(P-\hat {\pf})B_{T}}
G\rk_{{\ms E}}
 \right],
\end{align}
where $\hat {\pf}=d\Gamma(-i\nabla_k)$ and 
$\tilde \varrho_\eps(x)=\lk \e^{-\eps|\cdot|^2/2}\one_\la^\perp /\sqrt{\omega}\rk^\vee{}(x)$, and $\delta_s(x) =
\delta(x-s)$ is the one-dimensional Dirac delta distribution with mass on $s$.
\el
\proof
See \cite{hir07} and Section \ref{appb}.
\qed
\bc{positivity}\TTT{Positivity improving}
Let $P=0$. Then 
$e^{-TH_\eps(0)}$ is positivity improving. 
\ec
\proof
By Lemma \ref{FKF} we have 
\begin{align}
(F, \e^{-2TH_\eps(0)}G)
\label{fkf-1}
=  \Ebb \left[  
\lk 
\JJ_{-T}
e^{-i\hat {\pf}\cdot B_{-T}}
 F, 
\e^{-\phi_{\rm E}(\int_{-T}^T
 \delta_s\otimes \tilde\varrho_\eps(\cdot-B_s) ds)} \JJ_T
e^{-i\hat {\pf}\cdot B_{T}}
G\rk_{{\ms E}}
 \right].
\end{align}
Since $J_t$ and  $e^{i\hat {\pf} \cdot B_T}$ are positivity preserving, and $J_t^\ast J_s=e^{-|t-s|\hf}$ is positivity improving, 
we have 
$(F, \e^{-2TH_\eps(0)}G)\geq0$ for $F\geq0$ and $G\geq0$. We can also deduce that 
$(F, \e^{-2TH_\eps(0)}G)\neq0$ in the same way as \cite{hir07}. Then 
$(F, \e^{-2TH_\eps(0)}G)>0$ follows and it implies the statement of the lemma. 
\qed

\subsection{Convergence on Fock vacuum}
In order to prove Theorem \ref{main} 
we need two ingredients:
\bi
\item[(1)]
convergence \kak{main1} on the Fock vacuum,  
\item[(2)]
uniform lower bound of $\he$ with respect to $\eps$. 
\ei
Let $\one =\{1,0,0,\cdots,\}\in\fff$ be the Fock vacuum. 
In particular, for 
$F=\one =G$, we can see  the corollary  below. 
\bc{24}\TTT{Vacuum expectation}
It follows that 
\eq{2}
(\one , \e^{-2T\he }  \one ) = 
 \Ebb\!\left[
\e^{iP\cdot (B_T-B_{-T})}  \e^{\frac{g^2}{2} S_\eps } \right],
\en
where
\eq{regint}
S_\eps =  \int _{-T}^T ds\int_{-T}^T dt
 W_\eps (B_t-B_s,t-s)
\en
is the pair interaction given by the pair  potential $W_\eps: \BR \times\RR \to \RR$: 
\begin{equation}
\label{eq:nelsonWreg}
W_\eps (x,t) =  
\int_{|k|\geq \la} \frac{1}{2\omega(k)} \e^{-\eps |k|^2} \e^{-ik\cdot x} \e^{-\omega(k)|t|} dk.
\end{equation}
\ec
\proof
This follows directly from Lemma \ref{FKF}. 
\qed
It can be seen that the pair potential  $W_\eps (B_t-B_s,t-s)$ 
is singular at the diagonal part $t=s$.
We shall remove the diagonal part by using the It\^o formula. 
We  introduce 
the function
\begin{equation}
\label{eq:phiW}
\varrho_\eps(x,t) =  \int_{|k|\geq \la} \frac{\e^{-\eps |k|^2} \e^{-ik\cdot x-\omega(k)|t|}}{2 \omega(k)}
\beta(k) dk,\quad \eps\geq 0,
\end{equation}
where $\beta(k)$ is given by \kak{beta}, 
and it is shown 
by 
the It\^o formula  that
\begin{multline}
\label{eq:Wdec}
\int_{s}^{S} W_\eps (B_t-B_s,t-s) dt \\
= \varrho_\eps(0,0) -\varrho_\eps(B_{S}-B_s,
S-s) +
\int_{s}^{S} \nabla\varrho_\eps(B_t-B_s,t-s)\cdot dB.
\end{multline}
Here
$\varrho_\eps(0,0)$ can be regarded as the diagonal part of $W_\eps$ and turns to be a renormalization term, since 
$\varrho_\eps(0,0)\to -\infty$ as $\eps\to 0$.  
Let 
$$S_\eps^{\rm ren}=S_\eps-4T\varrho_\eps(0,0),\quad \eps>0,$$
which is represented as 
\begin{align}
S_\eps^{\rm ren}
=
S_\eps^{OD}&+2 \int_{-T}^T ds\lk 
\int_{s}^{[s+\tau]}\nabla\varrho_\eps(B_t-B_s,t-s) ds \rk 
\cdot  dB_t  \non\\
&-2 \int_{-T}^T \varrho_\eps(B_{[s+\tau]}-B_s,[s+\tau]-s)ds.
\end{align}
Here $0<\tau<T$ is an arbitrary number, 
$S_\eps^{OD}$ denotes the off-diagonal part which 
is given by 
$$S_\eps^{OD}=
2\int_{-T}^T ds\int_{[s+\tau]}^T W_\eps(B_t-B_s,t-s) dt$$
and $[t]=-T\vee t\wedge T$, 
and 
the integrand is  given by 
\begin{align*}
\nabla \varrho_\eps(X,t)=\int_{|k|\geq \la}
\frac{-ik e^{-ikX}e^{-|t|\omega(k)}e^{-\eps|k|^2}}{2\omega(k)}\beta(k)  dk.
\end{align*}
\begin{proposition}
\label{muri}
(1) 
It holds that 
\eq{mmm1}
\lim_{\eps\downarrow 0}
\Ebb \left[
 \left|
\e^{\frac{g^2}{2}S_\eps^{\rm ren} }-\e^{\frac{g^2}{2}S_0^{\rm ren} }\right|\right]=0.
\en
(2) 
There exists a constant $c>0$ such that 
for all $\eps\geq0 $,
\eq{mmm2}
\Ebb \left[
\e^{\frac{g^2}{2}S_\eps^{\rm ren} }
\right]
 \leq e^{Tc}
\en
\end{proposition}
Here 
$$
S_0^{\rm ren}=2 \int_{-T}^T ds\lk 
\int_{-T}^t \nabla\varrho_0(B_t-B_s,t-s) ds \rk 
\cdot  dB_t  
-2 \int_{-T}^T \varrho_0(B_{T}-B_s,T-s)ds.$$
\proof 
This can be proven by a minor modification of \cite[Section 2]{ghl13}.
\qed
From this proposition we can derive the lemma below immediately.
\bl{maindenai}
It follows that
\eq{sasa21}
\lim_{\eps\downarrow 0} (\one ,\e^{-2T(\he +g^2\varrho_\eps(0,0))}  \one )
=  \Ebb
\!\left[ \e^{iP\cdot (B_T-B_{-T})} \e^{\frac{g^2}{2} S_0^{\rm ren} }
\right].
\en
\el

\subsection{Existence of ground states for $P=0$}
\label{223}
We shall show a  uniform lower bound 
of $H_\eps(P)+g^2  \varrho_\eps(0,0)$ with respect to 
$\eps\geq 0$, 
and give the proof of Theorem \ref{main}. 
Let $E_\eps(P)=\is(H_\eps(P))$. 
\bc{dia}\TTT{Diamagnetic inequality}
Let $\eps>0$. Then  
$E_\eps(0)\leq E_\eps(P)$ follows for every  $P\in\BR$. 
\ec
\proof
By functional integral representation 
\kak{fkf} it follows that 
$$|(F, e^{-TH_\eps(P)}G)|\leq (|F|, e^{-TH_\eps(0)}|G|).$$
This yields the inequality  
$E_\eps(0)\leq E_\eps(P)$. 
\qed
Intuitively 
$$\grt=e^{-TH_\eps(0)}\one /\|e^{-TH_\eps(0)}\one \|$$ is a sequence converging to a ground state. 
Let $\gamma(T)=(\one , \grt)^2$, i.e., 
\eq{3-1}
\gamma(T)=\frac{(\one , e^{-TH_\eps(0)}\one )^2}{(\one , e^{-2TH_\eps(0)}\one )}.
\en
The useful lemma concerning the existence and absence of 
 ground states is the lemma below. 
\bl{ex1}
There exists a ground state of $H_\eps(0)$ 
if and only if $\d \lim_{T\to\infty}\gamma(T)>0$.
\el
\proof
This proof is taken from \cite{lms02}.
Suppose that $\is (H_\eps(0) )=0$ and set 
$\d \lim_{T\to\infty}\gamma(T)=a$. 
Suppose that  $a=0$ and the ground state $\gr$ exists. 
Then $\lim_{T\to\infty} e^{-TH_\eps(0)}=\one_{\{0\}}(H_\eps(0))$. 
Since $\gr>0$ by the fact that $e^{-TH_\eps(0)} $ is positivity improving,  
it follows that $a=(\one, \gr)>0$. It contradicts $a>0$. 
Thus the ground state does not exist. 
Next suppose $a>0$. 
Then $\sqrt{\gamma(T)}\geq\eps$ for sufficiently large $T$. Let $dE$ be the spectral measure of 
$H_\eps(0)$. 
Thus we have 
$$\sqrt{\gamma(T)}=
\frac{\int_0^\infty e^{-Tu  }dE}{(\int_0^\infty e^{-2Tu  } dE)^{1/2}}\leq
\frac{\int_0^\delta e^{-Tu  }dE+\int_\delta^\infty e^{-Tu  }dE}{(\int_0^\delta e^{-2Tu  } dE)^{1/2}}.$$
Then we can derive that 
$$\sqrt{\gamma(T)}\leq 
\frac{(\int_0^\delta e^{-2Tu  }dE)^\han
E([0,\delta])^\han+e^{-T\delta}}{(\int_0^\delta e^{-2Tu  } dE)^{1/2}}=
E([0,\delta])^\han+\frac{1}{(\int_0^\delta e^{-2T(u  -\delta)} dE)^{1/2}}.$$
Take $T\to\infty$ on both sides above, we have 
$\sqrt\eps\leq E([0,\delta])^\han$. Thus taking $\delta\downarrow 0$, we have  
$\sqrt\eps\leq E(\{0\})^\han$. Thus the ground state exists. 
\qed
Using the lemma above we can show the existence of the ground state of $H_\eps(0)$. 
\bl{ex2}
For all $\eps>0$, $H_\eps(0)$ has the ground state and it is unique.
\el
\proof
The uniqueness follows from the fact that 
$e^{-tH_\eps(0)}$ is positivity improving. 
It remains to show the existence of ground state, 
which is proven by using Lemma  \ref{ex1}.  
By the Feynman-Kac type formula we have 
$$\gamma(T)=\frac{\lk
\Ebb[e^{\frac{g^2}{2}\int_0^T dt\int_0^T ds
W_\eps}]\rk^2}{
\Ebb[e^{\frac{g^2}{2}\int_{-T}^T dt\int_{-T}^T dsW_\eps}]}.$$
By the reflection symmetry of the Brownian motion we see that 
$$\gamma(T)=\frac{\Ebb[e^{\frac{g^2}{2}\int_0^T dt\int_0^T dsW_\eps}]\Ebb[e^{\frac{g^2}{2}\int_{-T}^0 dt\int_{-T}^0 dsW_\eps}]}{
\Ebb[e^{\frac{g^2}{2}\int_{-T}^T dt\int_{-T}^T dsW_\eps}]}$$ and also the Markov property yields that 
$$\gamma(T)=\frac{\Ebb[e^{\frac{g^2}{2}\int_0^T dt\int_0^T dsW_\eps+
\frac{g^2}{2}\int_{-T}^0 dt\int_{-T}^0 dsW_\eps
}]}{
\Ebb[e^{\frac{g^2}{2}\int_{-T}^T dt\int_{-T}^T dsW_\eps}]}.$$
Then we obtain that 
$$\gamma(T)=\frac{\Ebb[e^{\frac{g^2}{2}
\int_{-T}^T dt\int_{-T}^T dsW_\eps-g^2\int_{-T}^0dt\int_0^T W_\eps}]}
{
\Ebb[e^{\frac{g^2}{2}\int_{-T}^T dt\int_{-T}^T dsW_\eps}]}.$$
Notice that 
$$\int_{-T}^0dt\int_0^T ds W_\eps
\leq 
\int_\BR \kp \frac {e^{-\eps|k|^2}}{\omega(k)^3}
\lk1-e^{-\omega(k) T}\rk^3 dk\leq 
\int_\BR \kp \frac{ e^{-\eps|k|^2}}{\omega(k)^3}dk
.$$
Hence we conclude that 
\eq{big}
\gamma(T)\geq 
\exp\lk -g^2\int_\BR \kp \frac{ e^{-\eps|k|^2}}{\omega(k)^3}dk
\rk >0
\en
for all $T>0$.
Then the lemma follows. 
\qed
\subsection{Uniform lower bounds and the proof of main theorem}
In this section we show a  uniform lower bound of the bottom of the spectrum of $\he +g^2 \varrho_\eps(0,0)$
with respect to $\eps>0$.
Thanks to the diamagnetic inequality, the estimate of the uniform lower bound for any $P$ can be reduced to that of $P=0$. We note that the diamagnetic inequality 
$E(0)\leq E(P)$ can be derived through a functional integration in Corollary \ref{dia}.

\bl{nelson}
There exists $C \in \RR$ such that 
$\he -g^2 \varrho_\eps(0,0)>-C$, uniformly in $\eps>0$.
\el
\proof
Let $\gr$ be the ground state of $H_\eps(0)$. 
Since $e^{-tH_\eps(0)}$ is positivity improving, 
we see that  
 $(\one , \gr)\ne 0$ and 
then 
 $$\d E_\eps(0)-g^2\varrho_\eps(0,0)=-\lim_{T\to\infty}\frac{1}{T} \log
 (\one , e^{-T(H_\eps(0)-g^2\varrho_\eps(0,0))}\one )>-C$$
by Proposition \ref{muri},  
 where $C$ is independent of $\eps > 0$.
 By the diamagnetic inequality 
 $E_\eps(0)\leq E_\eps(P)$ we then derive that 
 $$E_\eps(P)- g^2\varrho_\eps(0,0)\geq-C.$$
 Then the lemma follows. 
\qed

Now we extend the result  from 
Fock vacuum $\one $ 
to more general
vectors of the form $F(\phi(f_1),\ldots,\phi(f_n))$, with $F\in \ms  S(\RR^n)$, where
$\phi(f)$ stands for a scalar field given by
$$\phi(f)=\frac{1}{\sqrt 2} (\add( {\hat f}/\sqrt\omega )+a(\widetilde{\hat f}/\sqrt\omega)).$$
Consider the subspace 
$$
\D= \lkk  F(\phi(f_1),\ldots,\phi(f_n)) 
| F\in \ms S(\RR^n), f_j\in H_{-\han}(\BR), j=1,...,n, n\geq 1\rkk,$$
which is dense in $\fff$.

\bl{FKF2}
(1) Let $ \rho_j\in H_{-\han}(\BR)$ for $j=1,2$, and $\alpha,\beta\in\CC$. Then
\begin{eqnarray}
\lim_{\eps\downarrow0} 
( \e^{\alpha\phi(\rho_1)},
\e^{-2T(\he +g^2\varrho_\eps(0,0))}   
\e^{\beta\phi(\rho_2)}) 
\label{main22}
=  \Ebb\!\left[\e^{iP\cdot (B_T-B_{-T})}
  \e^{\frac{g^2}{2}
S_0^{\rm ren} +\frac{1}{4}\xi} \right],
\end{eqnarray}
where
\begin{align*}
\xi=\xi(g)
&=
\bar \alpha^2\|\rho_1/\sqrt\omega \|^2+\beta^2\|\rho_2/\sqrt\omega\|^2+2\bar
\alpha \beta (\rho_1/\sqrt\omega, \e^{-2T\omega} \rho_2/\sqrt\omega) \\
&
\qquad + 2\bar \alpha g 
\int_{-T}^Tds \int_\BR dk \frac{\hat  \rho_1(k)}{\sqrt{\omega(k)}}\kp
\e^{-|s-T|\omega(k)}\e^{-ikB_s} \\
&
\qquad
+ 2\beta  g
\int_{-T}^Tds \int_\BR dk \frac{\hat  \rho_2(k)}{\sqrt{\omega(k)}}\kp
\e^{-|s+T|\omega(k)}\e^{-ikB_s}.
\end{align*}
(2)
Let $\Phi=F(\phi(u_1),\ldots,\phi(u_n))$ and  $\Psi= G(\phi(v_1),\ldots,\phi(v_m)) \in \D$.
Then
\begin{align}
&\lim_{\eps\downarrow0}
(\Phi, \e^{-2T (\he +g^2 \varrho_\eps(0,0))} \Psi)\non \\
&=
(2\pi)^{-(n+m)/2} \int_{\RR^{n+m}}dK_1 dK_2 \ov{\hat  F(K_1)} \hat  G(K_2) 
   \Ebb\!\left[ \e^{iP\cdot (B_T-B_{-T})}  \e^{\frac{g^2}{2}
S_0^{\rm ren} + \frac{1}{4} \xi(K_1,K_2)} \right],
\label{310}
\end{align}
where
\begin{align*}
\xi(K_1,K_2)
&=
-\|K_1\cdot u/\sqrt\omega\|^2 -\|K_2\cdot v /\sqrt\omega\|^2 -
2 (K_1 \cdot u/\sqrt\omega, \e^{-2T\omega} K_2 \cdot v/\sqrt\omega)\\
& \qquad
- 2i g  \int_{-T}^Tds \int_\BR dk
\frac{K_1\cdot \hat  u(k)}{\sqrt{\omega(k)}}\kp \e^{-|s-T|\omega(k)}\e^{-ikB_s} \\
&
\qquad + 2i g  \int_{-T}^Tds \int_\BR dk
\frac{K_2\cdot \hat  v(k)}{\sqrt{\omega(k)}}\kp \e^{-|s+T|\omega(k)}\e^{-ikB_s}
\end{align*}
and $u=(u_1,...,u_n)$, $v=(v_1,...,v_m)$.
\el
\proof
(1) follows from Lemma \ref{FKF}. 
(2)  follows from 
$$\Phi=F(\phi(u_1),\ldots,\phi(u_n))
=(2\pi)^{-n/2}\int_{\RR^n} 
\hat F(k_1,\cdots,k_n) e^{i\sum_{j=1}^n k_j \phi(u_j)} dk_1\cdots dk_n$$
and 
Lemma  \ref{maindenai}.
\qed

Now we can complete the proof of the main theorem.

\medskip
\noindent
\emph{Proof of Theorem \ref{main}.}
Let $F, G\in \hhh$ and $C_\eps(F,G)=(F, 
\e^{-t(H_\eps(P)+g^2\varrho_\eps(0,0))}G)$. 
By Lemma  \ref{FKF} we
obtain that $C_\eps(F,G)$ is convergent as $\eps\downarrow0$, for every $F,G\in\D$. 
Since $\D$ is dense in $\hhh$, 
by the uniform bound
$
\|\e^{-t(H_\eps(P)+g^2\varrho_\eps(0,0))}\|
<\e^{tC}
$
obtained by Lemma   \ref{nelson} we can see that 
$\{C_\eps(F,G)\}_\eps$ converges 
also  for all $F,G\in\hhh$ by a simple approximation. 
Let $C_0(F,G)=\lim_{\eps\downarrow 0}C_\eps(F,G)$.
Hence  
$$|C_0(F,G)|\leq \e^{tC}\|F\|\|G\|,$$
and 
 there exists a bounded operator
$T_t$ such that
$$
C_0(F,G)=(F, T_t G),\quad 
F,G\in \hhh$$ by 
the  Riesz theorem.
Thus $\slim_{\eps\downarrow 0}\e^{-t(\he +g^2\varrho_\eps(0,0))}=T_t$ follows. Furthermore, we also see 
that 
$$\d 
\slim_{\eps\downarrow 0} \e^{-t(\he +g^2\varrho_\eps(0,0))}
\e^{-s(\he +g^2\varrho_\eps(0,0))}=\slim_{\eps\downarrow 0}
\e^{-(t+s)(\he +g^2\varrho_\eps(0,0))}=T_{t+s}.$$
Since the left-hand side above is $T_t T_s$, the semigroup property of $T_t$ follows. Since
$\e^{-t(\he +g^2\varrho_\eps(0,0))}$ is a symmetric semigroup, $T_t$ is also symmetric. Moreover by
the functional integral representation \kak{310} the functional $(F, T_t G)$ is continuous at $t=0$
for every $F, G \in \D$. 
Since $\D$ is dense in $\hhh$ and $\|T_t\|$ is uniformly bounded, 
it also follows that  $T_t$ is strongly continuous at $t=0$. 
Then $T_t$, $t\geq0$,  is strongly continuous one-parameter symmetric semigroup. 
Thus the
semigroup version of Stone's theorem 
\cite[Proposition 3.26]{lhb11} implies that 
there exists a
self-adjoint operator $\hni$, bounded from below, such that
$$T_t=\e^{-t\hni},\quad t\geq0.$$
Hence the proof is
completed by setting $E_\eps=-g^2\varrho_\eps(0,0)$.
\qed
Let $E_{\rm ren}(P)=\is(H_{{\rm ren}}(P))$.
\bc{d-m}\TTT{Diamagnetic inequality}
It holds that $E_{\rm ren}(0)\leq E_{\rm ren}(P)$.
\ec
\proof
From 
inequality $|(F, e^{-T(H_\eps(P)-E_\eps)}G)|\leq 
(|F|, e^{-T(H_\eps(0)-E_\eps)}|G|)$ it follows that 
$|(F, e^{-TH_{{\rm ren}}(P)}G)|\leq 
(|F|, e^{-TH_{{\rm ren}}(0)}|G|)$. Then the corollary follows.
\qed

\section{Existence of renormalized ground state for $d=2$}
Let us suppose $$d=2.$$
In the case of $d=2$ we can procedure the renormalization similar to the case of $d=3$. 
The renormalization is however not needed in the case of $d=2$, since 
$\varrho_\eps(0,0)$ converges to the finite number $\varrho_0(0,0)$ as 
$\eps\to 0$. 
One important conclusion 
of Theorem \ref{main} is the existence of 
a ground state of $H_{\rm ren}(0)$ for $d=2$.
\bl{big2}
It follows that 
\eq{big3}
\gamma(T)=
\frac{(\one , e^{-TH_{\rm ren}(0)}\one )^2}{(\one , e^{-2TH_{\rm ren}(0)}\one )}
>\exp\lk -g^2\int_{\RR^2} \kp \frac{ 1}{\omega(k)^3}dk
\rk >0
\en
\el
\proof
By \kak{big} we have 
\begin{align*}
\gamma(T)=
\frac{(\one , e^{-TH_\eps(0)}\one )^2}{(\one , e^{-2TH_\eps(0)}\one )}
\geq 
\exp\lk -g^2\int_{\RR^2} \kp \frac{e^{-\eps|k|^2}}{\omega(k)^3}dk
\rk >0.
\end{align*}
Take the limit of $T\to\infty$ on both sides we can derive \kak{big3}. 
\qed
\bt{main2}
\TTT{Existence of the ground state}
For arbitrary values of $g$, $H_{\rm ren}(0)$ 
has a ground state $\gre$ such that $(\one , \gre)\ne 0$. 
\et
\proof
By Lemma  \ref{big3}  
we have 
\begin{align}\label{big5}
\lim_{T\to\infty} 
\frac{(\one , e^{-TH_{\rm ren}(0)}\one )^2}{(\one , e^{-2TH_{\rm ren}(0)}\one )}
>
\exp\lk -g^2\int_{\RR^2} \kp \frac{1}{\omega(k)^3}dk
\rk >0.
\end{align}
On the other hand 
we see that 
$$\lim_{T\to\infty} 
\frac{(\one , e^{-TH_{\rm ren}(0)}\one )^2}{(\one , e^{-2TH_{\rm ren}(0)}\one )}
=\|P_g\one \|^2,$$
where $P_g$ denotes the projection to 
the subspace ${\rm Ker} (H_{\rm ren}-\is(H_{\rm ren}))$.
By \kak{big5} we derive that 
$\|P_g\one \|^2>0$, which implies 
$H_{\rm ren} $ has a ground state $\gre$ 
such that  $(\one , \gre)\ne0$.  
\qed

\section{Polaron model}
We introduce the polaron model in this section. The polaron model is similar to $\he$, and the UV renormalization can be seen in a similar manner to the Nelson model. 
The polaron Hamiltonian is defined by 
\eq{p1}
H^{\rm pol}(P)=\half (P-\pf)^2+N+g\Phi,\quad P\in\BR, 
\en
where 
$N$ denotes the number operator and 
$$\Phi=\frac{1}{\sqrt2}
\lk 
\add(\vp/\omega)+a(\tilde{\vp}/\omega)\rk.$$
Note that the test function is $\vp/\omega$ which is different from the test function $\vp/\sqrt\omega$ of the Nelson Hamiltonian. 
We discuss UV renormalization of the polaron model. 
The discussion is however easier than that of 
the Nelson model.
Let $\vp(k)=e^{-\eps|\k|^2/2}$, 
and $H^{\rm pol}(P)$ with $\vp(k)=e^{-\eps|\k|^2/2}$
is denoted by $H^{\rm pol}_\eps(P)$. 
The vacuum expectation of $e^{-T H^{\rm pol}_\eps(P)}$ is given by 
\eq{p2}
(\one , \e^{-TH^{\rm pol}_\eps(P)}  \one ) = 
 \Ebb\!\left[
\e^{iP\cdot B_T}  \e^{\frac{g^2}{2} S^{\rm pol}_\eps } \right],
\en
where
\eq{pregint}
S^{\rm pol}_\eps =  \int _0^T ds\int_0^T dt
 W_\eps^{\rm pol} (B_t-B_s,t-s)
\en
is the pair interaction for the polaron model and 
the pair  potential is given by 
\eq
{peq:nelsonWreg}
W_\eps^{\rm pol} (x,t) =  
\int_{|k|\geq \la} \frac{1}{2\omega(k)^2} \e^{-\eps |k|^2} \e^{-ik\cdot x} \e^{-|t|} dk.
\en
We can see that  
$$
W_\eps^{\rm pol} (x,t) =  
\frac{2\pi}{|x|}\int_{\la |x|}^\infty e^{-\eps u} \frac{\sin u}{u} du e^{-|t|}.$$
Let 
$$W_0^{\rm pol} (x,t) =  
\int_{|k|\geq \la} \frac{1}{2\omega(k)^2} 
\e^{-ik\cdot x} \e^{-|t|} dk$$
and we see that 
$W_\eps^{\rm pol}(x,t)\to W_0^{\rm pol}(x,t)$ for each $(x,t)$ as $\eps\downarrow 0$. 
Then it holds that 
\eq{pmmm1}
\lim_{\eps\downarrow 0}
\Ebb \left[
 \left|
\e^{\frac{g^2}{2}S^{\rm pol}_\eps }-\e^{\frac{g^2}{2}S^{\rm pol}_0 }\right|\right]=0.
\en
From this  we can prove the lemma below immediately.
Note that any renormalization is not needed. 
\bl{pmaindenai}
It follows that
\eq{psasa21}
\lim_{\eps\downarrow 0} (\one ,\e^{-TH^{\rm pol}_\eps(P)}  \one )
=  \Ebb
\!\left[ \e^{iP\cdot B_T} \e^{\frac{g^2}{2} S^{\rm pol}_0 }
\right].
\en
\el
Hence the theorem below is proven in the same way as the Nelson model. 
\bt{pmain}
\TTT{UV renormalization}
Let $P\in \BR$. 
Then 
there exists a self-adjoint operator $H^{\rm pol}_0(P)$ such that
\eq{pmain1}
\slim_{\eps\downarrow0}
\e^{-TH^{\rm pol}_\eps(P) }=\e^{-TH^{\rm pol}_0(P)},\quad T\geq0.
\en
\et

\bc{polaron}\TTT{Removal of infrared cutoff}
It follows that 
\eq{re}
\lim_{\la\to0} (\one, e^{-TH_0^{\rm pol}(P)}\one) =
\Ebb\left[e^{iP\cdot B_T}
e^{g^2\frac{\pi^2}{2}
\int_0^Tdt\int_0^Tds \frac{e^{-|t-s|}}{|B_t-B_s|}}\right].
\en
\ec
\proof
It can be seen that 
$$W_0^{\rm pol}(x,t)=
\int_{|k|\geq \la}\frac{1}{2\omega(k)}e^{-ik\cdot x}e^{-|t|}dk\leq 
\frac{\pi^2+\delta }{|x|}e^{-|t|}$$
with some constant $\delta$, 
and 
$$\lim_{\la\to0}W_0^{\rm pol}(x,t)=
\frac{\pi^2 }{|x|}e^{-|t|}$$
for each $x$. 
It can be also checked that 
$\Ebb\left[
e^{g^2\frac{\pi^2}{2}
\int_0^Tdt\int_0^Tds \frac{e^{-|t-s|}}{|B_t-B_s|}}\right]$ is finite in the lemma below. 
Then the Lebesgue dominated convergence theorem yields the corollary. 
\qed
\bl{finite}
$\Ebb\left[
e^{g^2\frac{\pi^2}{2}
\int_0^Tdt\int_0^Tds \frac{e^{-|t-s|}}{|B_t-B_s|}}\right]$is finite. 
\el
\proof
We separate $[0,T]\times [0,T]$ into 
two regions as  
\begin{align*}
\int_0^Tdt\int_0^Tds =
\int_0^Tdt\int_t^Tds +
\int_0^Tdt\int_0^t ds.
\end{align*}
By the Schwarz inequality we have 
\begin{align}
&
\Ebb\left[
e^{g^2\frac{\pi^2}{2}\int_0^Tdt\int_0^Tds 
\frac{e^{-|t-s|}}{|B_t-B_s|}}\right]\non \\
&
\leq 
\lk 
\Ebb\left[
e^{2g^2\frac{\pi^2}{2}\int_0^Tdt\int_t^Tds 
\frac{e^{-|t-s|}}{|B_t-B_s|}}\right]\rk^\han 
\lk
\Ebb\left[
e^{2g^2\frac{\pi^2}{2}\int_0^Tdt\int_0^t ds 
\frac{e^{-|t-s|}}{|B_t-B_s|}}\right]\rk^\han\non \\
&\label{both}=\lk 
\Ebb\left[
e^{2g^2\frac{\pi^2}{2}\int_0^Tdt\int_t^Tds 
\frac{e^{-|t-s|}}{|B_t-B_s|}}\right]\rk^\han 
\lk
\Ebb\left[
e^{2g^2\frac{\pi^2}{2}\int_0^Tds
\int_s^T dt
\frac{e^{-|t-s|}}{|B_t-B_s|}}\right]\rk^\han. 
\end{align}
We estimate both sides of \kak{both}. 
By Jensen's inequality we have 
\begin{align*}
\Ebb\left[
e^{g^2 {\pi^2}\int_0^Tdt\int_t^Tds 
\frac{e^{-|t-s|}}{|B_t-B_s|}}\right]
\leq
\int_0^T \frac{dt}{T}
\Ebb\left[e^{g^2\frac{\pi^2}{2}\int_t^Tdt T  
\frac{e^{-|t-s|}}{|B_t-B_s|}}\right]. 
\end{align*}
We estimate
$\Ebb\left[e^{g^2{\pi^2}\int_t^Tdt T  
\frac{e^{-|t-s|}}{|B_t-B_s|}}\right]$. 
Let $\pro {\ms F}$ be the natural filtration of the Brownian motion $\pro B$. 
We can see  that 
\begin{align*}
&\Ebb\left[e^{g^2{\pi^2}\int_t^Tds T  \frac{e^{-|t-s|}}{|B_t-B_s|}}\right]=
\Ebb\left[
\Ebb\left[e^{g^2{\pi^2}\int_t^Tds T  \frac{e^{-|t-s|}}{|B_t-B_s|}}|\fff_t\right]\right]\\
&=
\Ebb\left[
\Ebb^{B_t}
\left[e^{g^2{\pi^2}\int_t^Tds T  \frac{e^{-|t-s|}}{|
B_{0}-B_{s-t}|}}\right]\right]
=
\int _\BR dy 
(2\pi t)^{-3/2} e^{-|y|^2/(2t)}
\Ebb^{y}
\left[e^{g^2{\pi^2}\int_t^Tds T  
\frac{e^{-|t-s|}}{|
B_0-B_{s-t}|}}\right]\\
&
=\int _\BR dy 
(2\pi t)^{-3/2} e^{-|y|^2/(2t)}
\Ebb^{y}
\left[e^{g^2{\pi^2}\int_t^Tds T  
\frac{e^{-|t-s|}}{|
B_{s-t}-y|}}\right].
\end{align*}
Since the potential 
$V(x)= |x|^{-1}$  is a Kato class potential, we have  
\begin{align*}
\sup_y \Ebb^{y}
\left[e^{g^2{\pi^2}\int_t^Tds T  \frac{e^{-|t-s|}}{|
B_{s-t}-y|}}\right]\leq e^{a(T-t)}
\end{align*}
with some $a$. 
Hence 
$\Ebb\left[
e^{g^2 {\pi^2}\int_0^Tdt\int_t^Tds 
\frac{e^{-|t-s|}}{|B_t-B_s|}}\right]<\infty$. 
Similarly it can be shown that 
$\Ebb\left[
e^{g^2 {\pi^2}\int_0^Tds\int_s^Tdt 
\frac{e^{-|t-s|}}{|B_t-B_s|}}\right]<\infty$ and 
hence \kak{both} is finite. 
\qed

\appendix

\section{Schr\"odinger representation and Euclidean field}
\label{appb}
In this section  Hilbert spaces $H_{-\han}(\RR^3)$ and 
$H_{-1}(\RR^4)$
are  given by \kak{yoshida}.
It is well known that the boson Fock space $\ffff $ is unitarily equivalent to $L^2(Q,\mu)$, where this
space consists of square integrable functions on a probability space $(Q,\Sigma,\mu)$. Consider the
family of Gaussian random variables 
$\{\phi (f),  f\in H_{-1/2}(\BR)\}$ on $(Q,\Sigma,\mu)$ such
that $\phi (f)$ is linear in 
$f\in H_{-\han}(\RR^3)$,  and their mean and covariance are given by
$$
\Ebb_{\mu}[\phi (f)]=0 \quad \mbox{and} \quad \Ebb_{\mu}[\phi (f)\phi (g)]
=\half (f, g)_{H_{-\han}(\RR^3)}.
$$
Given this space, the Fock vacuum $\one_\ffff $ is unitary equivalent to $\one_{L^2(Q)}\in L^2(Q)$, and
the scalar field $\phi(f)$ is unitary equivalent to $\phi (f)$ as operators, i.e., $\phi (f)$ is
regarded as multiplication by $\phi (f)$. Then the linear hull of the vectors given by the Wick products
$:\prod_{j=1}^n \phi (f_j):$ is dense in $L^2(Q)$, where recall that Wick product is recursively defined
by
\begin{eqnarray*}
&&
{:\phi (f):} \; =\phi (f)  \\
&&
{:\phi (f)\prod_{j=1}^n \phi (f_j):} \: =
\phi (f):\prod_{j=1}^n \phi (f_j):-\half\sum_{i=1}^n (f,f_i)_{H_{-\han}(\RR^3)} :\prod_{j\ne i}^n \phi (f_j):
\end{eqnarray*}
This allows to identify $\ffff $ and $L^2(Q)$, which we have done in \kak{fkf}, i.e., $F\in \hhh$ can be regarded
as a function $\RR^{3N}\ni x\mapsto F(x)\in L^2(Q)$ such that $\int_{\RR^{3N}}\|F(x)\|^2_{L^2(Q)} dx<\infty$.

To construct a Feynman-Kac type formula  we use a Euclidean field. Consider the family of Gaussian random
variables $\{\phi_{\rm E}(F),  F\in H_{-1}(\RR^4)\}$ with mean and covariance
$$
\Ebb_{\mu_E}[\phi_E(F)]=0 \quad \mbox{and} \quad \Ebb_{\mu_E}[\phi_E(F)\phi_E(G)]=\half (F, G)_{H_{-1}(\RR^4)}
$$
on a chosen probability space $(Q_E, \Sigma_E, \mu_E)$. Note that for $f\in H_{-1/2}(\BR)$ the relations
$
\delta_t\otimes f\in H_{-1}(\RR^4) \quad  \mbox{and} \quad \|\delta_t\otimes f\|_{H_{-1}(\RR^4)} =
\|f\|_{H_{-1/2}(\BR)}
$
hold, where $\delta_t(x) = \delta(x-t)$ is Dirac delta distribution with mass on $t$. The family of identities
used in \kak{fkf}
is  then given by $\JJ_t: L^2(Q)\to {\ms E}$, $t\in\RR$, defined by the relations
$$
\JJ_t \one_{L^2(Q)}=\one_{{\ms E}}$$
and 
$$
\JJ_t {:\prod_{j=1}^m \phi(f_j):} \;=\; {:\prod_{j=1}^m \phi_{\rm E}(\delta_t\otimes f_j):}.$$
Under the identification 
$\ffff \cong L^2(Q)$ it follows that 
$$(\JJ_t F, \JJ_s G)_{{\ms E}}=
(F, \e^{-|t-s|\hf }G)_\ffff $$ for $F, G\in\ffff $.

\section{The Nelson model}
\label{A}
The Nelson Hamiltonian $H$ is a self-adjoint operator acting in the Hilbert space 
$$\LR\otimes\fff\cong\int_\BR^\oplus \fff dx,$$
which is given by 
\eq{nelson1}
H=(-\half \Delta+V)\otimes \one+\one\otimes\hf+g \int_\BR^\oplus \phi(x) dx,
\en
where 
the interaction is defined by 
$$\phi(x)
  =  
\frac{1}{\sqrt 2}\lk
\add(\vp/\sqrt{\omega}e^{i(\cdot,x)})+
a(\widetilde{\vp}/\sqrt{\omega}e^{-i(\cdot,x)})
\rk.$$
$H$ is self-adjoint on $D(\hp)\cap D(\hf)$.
A point charge limit of $H$,  
$\vp (k) \to \one $, 
is studied in \cite{nel64,nel64b} and recently 
in \cite{ghps12, ghl13}. It is also shown in \cite{hhs05} that 
a point charge limit of $H$ has a ground state. 


We see the relationship between $H$ and $H(P)$. 
The total momentum $P_{{\rm tot},\mu}$ is defined by 
$P_{{\rm tot},\mu}=
-i\nabla_\mu\otimes\one+\one\otimes\pf_\mu$, $\mu=1,2,3$.
Let $V=0$ in $H$.  
Then $H$ becomes a translation invariant operator, which implies that 
$$[H, P_{{\rm tot},\mu}]=0,\quad 
\mu=1,2,3.$$
Thus 
$H$ can be decomposed with respect 
to the spectrum of total momentum 
 $P_{{\rm tot},\mu}$
  and it is known that 
 \eq{decp}
H\cong \int_\BR^\oplus H(P) dP.
\en

\noindent
{\bf Acknowledgments:}
The author acknowledges support of Challenging Exploratory Research 15K13445 from JSPS, 
and thanks for the kind hospitality of 
51 Winter School of Theoretical Physics
Ladek Zdroj, Poland, 9 - 14 February 2015. 
Moreover he also thanks Tadahiro Miyao who 
gives an idea to solve Lemma \ref{nelson} which is 
a key ingredient in this paper.

{

}


\begin{thebibliography}{99}


\bibitem[GHPS12]{ghps12}
C. G\'erard, F. Hiroshima, A. Panati and A. Suzuki,
 Removal of the UV cutoff for the Nelson model with variable coefficients,
 {\it Lett Math Phys}, {\bf 101} (2012), 305--322.


\bibitem[GHL13]{ghl13} M. Gubinelli, F. Hiroshima and J. Lorinczi, Ultraviolet renormalization of the Nelson Hamiltonian through functional integration, {\it J. Funct. Anal.} {\bf  267} (2014), 3125--3153.


\bibitem[HHS05]{hhs05}
M. Hirokawa, F. Hiroshima and H. Spohn, 
Ground state for point particle interacting through a massless scalar Bose field, {\it Adv. in Math.}{\bf  191} (2005), 339--392. 

\bibitem[Hir07]{hir07} F. Hiroshima, Fiber Hamiltonians in nonrelativistic quantum electrodynamics, {\it J. Funct. Anal.} {\bf 252} (2007) 314--355.


\bibitem[LHB11]{lhb11}J. L\H orinczi, F. Hiroshima and V. Betz, 
{\it Feynman-Kac-Type Theorems and
Gibbs Measures on Path Space},  
Studies in Mathematics {\bf 34}, de Gruyter, 2011.

\bibitem[LMS02]{lms02}J. L\H orinczi, R.A.Minlos and H. Spohn, 
The infrared behavior in Nelson's model of quamtum particle coupled to a massless scalar field, 
{\it Ann. Henri Poincar\'e} {\bf 3} (2002), 1--28.


\bibitem[Nel64a]{nel64}
E. Nelson,
Interaction of nonrelativistic particles with a quantized scalar field,
{\it J.   Math.   Phys. } {\bf  5}  (1964),   1190--1197.


\bibitem[Nel64b]{nel64b}
E. Nelson,
Schr\"odinger particles interacting with a quantized scalar field, in: {\it Proc. Conference
on Analysis in Function Space}, W. T.  Martin and I. Segal (eds.), p. 87, MIT Press, 1964.


\end{thebibliography}
\end{document}